\documentclass[preprint,12pt]{elsarticle}

\usepackage{amssymb}
\usepackage{graphics}
\usepackage{amsmath}
\usepackage{color}

\journal{Journal of Molecular Liquids}

\begin{document}

\begin{frontmatter}



\title{Fluid-fluid phase behaviour in the explicit solvent ionic model: hard spherocylinder solvent molecules}


\author{M.~Hvozd, T.~Patsahan, O.~Patsahan, M.~Holovko}

\address{Institute for Condensed Matter Physics of the National
Academy of Sciences of Ukraine, 1 Svientsitskii St., 79011 Lviv,
Ukraine}

\begin{abstract}
We study a fluid-fluid phase transition of the explicit solvent model represented as a mixture of
the restricted primitive model (RPM) of ionic fluid and neutral hard spherocylinders (HSC). 
To this end, we combine two theoretical approaches, i.e., the scale particle theory (SPT) and the associative 
mean spherical approximation (AMSA). Whereas the SPT is sufficient to provide a rather good description of 
a reference system  taking into account hard-core interactions,
the AMSA is known to be efficient  in treating the Coulomb interactions between the ions. 
Alternatively, we also use the mean spherical approximation (MSA) for  comparison.
In general,  both approximations lead to similar qualitative results for the phase diagrams:
 the region of coexisting envelope becomes broader and shifts towards larger densities 
and higher temperatures when the pressure increases. However, the AMSA and the MSA produce different concentration dependences, 
i.e., contrary to the MSA, the AMSA phase diagrams show that the high-density phase mostly consists of the ions for all  pressures considered.
To demonstrate the effect of asphericity of solvent molecules on the fluid-fluid phase transition, we consider an ``equivalent''  mixture 
in which the HSC particles are replaced by hard spheres (HS) of the same volume. 
It is observed that in the case of HSC solvent (RPM-HSC model), the region of phase coexistence is wider than for the case of the solvent molecules 
being of spherical shape (RPM-HS model).  It is also found that the critical temperature is higher in the RPM-HSC model than in the RPM-HS model,
though it becomes the same at higher pressures in the MSA, while in the AMSA this difference remains essential.
\end{abstract}


\end{frontmatter}

\section{Introduction}

The description of the thermodynamic behaviour  of ionic solutions including  their phase separation is important in many fields such as biology, electrochemistry, and material science.
The simplest model capable of capturing the main features of the  systems with electrostatic interactions is
the restricted primitive model (RPM).  In this model, the ionic fluid is
modelled as an electroneutral binary mixture of charged hard spheres (HS) of equal diameter and valency immersed in a structureless dielectric continuum.  A major drawback of this model is the use of the solvent continuum assumption in which the effects of solvent structure are totally neglected.
The simplest possible model which treats the solvent explicitly is a mixture of the RPM fluid and neutral hard spheres (the RPM-HS mixture).  
In such a model, the polar nature of the solvent is represented implicitly by a continuum
background with a dielectric constant. The RPM-HS mixture undergoes a demixing phase transition in the phases of different ion concentrations. 
The phase behaviour of the RPM-HS mixture was theoretically studied by using the mean-spherical approximation (MSA) \cite{Kenkare_SPM} and the 
pairing MSA (PMSA) \cite{Zhou_SPM}. More recently \cite{Patsahan-Patsahan:2018}, a comparative study of the fluid-fluid phase separation in the RPM-HS 
mixture has been performed using three theoretical approaches:  the random phase approximation (RPA) with the Weaks-Chandler-Anderson  regularization of 
the Coulomb potential  (the WCA approximation) \cite{cha,wcha}, the mean spherical approximation (MSA) \cite{Waisman72,WaismanLeb72,Blum75,Blum74}, and  
the  associative mean spherical approximation (AMSA)  \cite{HolKalyuzh91,Hol05}. The results have demonstrated that the AMSA  leads to the best agreement 
with the available simulation data \cite{Kristof_SPM} when the  association constant  proposed by Olaussen and Stell is used \cite{Olaussen91}. 
 
In this paper, we study the fluid-fluid phase diagram of an ionic solution model in which the solvent molecules are of non-spherical shape.
More precisely, we focus on the mixture of the RPM and uncharged hard spherocylinders (HSC) and refer to this system as the RPM-HSC mixture.  
In order to clarify the effect of asphericity of solvent molecules on the fluid-fluid phase transitions, we also consider a mixture 
of RPM fluid and ''equivalent'' hard spheres, which are taken of the same volumes as HSC particles in the previous case, and refer to this system
as the RPM-HS mixture.
Therefore, we investigate two similar models, but in the first model (RPM-HSC) neutral solvent molecules are of elongated shape like 
spherocylinders and in the second model (RPM-HS) neutral solvent molecules are spherical. 

For both the RPM-HSC and RPM-HS models, we use the AMSA theory to evaluate the contribution to thermodynamic properties due to
electrostatic interactions of the ionic component.
According to this theory the ions are considered as a system of free ions and ion pairs which are in a chemical equilibrium
defined by the mass action law (MAL)~\cite{HolKalyuzh91,Hol05,Blum95,Bernard96}. The AMSA theory reduces to the MSA when the association phenomena
between ions is neglected, i.e., ion pairing is not taken into account in the case of MSA. 
In our study we provide a comparison of the phase diagrams obtained from both the AMSA and MSA theories.

To evaluate the contribution of hard-core interactions to the thermodynamics of the RPM-HSC model, a reliable description of 
the binary system consisting of neutral HS and HSC particles (HS-HSC mixture) is necessary. 
For this purpose we apply the scale particle theory (SPT) originally developed by Reiss and coworkers  \cite{Reiss:1959,Reiss:1960}
and then extended to anisotropic particles in \cite{Gibbons:1969,cotter1970statistical,cotter1970,cotter1974hard}.
Recently \cite{HolovkoShmot2014,HolovkoHvozd2017}, the SPT has been developed to the case of a HS-HSC mixture. 
In \cite{HolovkoHvozd2017}, analytical expressions for the free energy, pressure and chemical potentials
are derived and, on this ground, an isotropic-nematic phase transition in the HS-HSC mixture is studied. 
Furthermore, the SPT theory was extended to the description of thermodynamic
properties and orientation ordering of the HS-HSC mixture confined in a disordered porous medium \cite{HvozdPatsahanHolovko:2018},
where the Carnahan-Starling-like correction (CS) and Parsons-Lee (PL) corrections \cite{Parsons1979,Lee1987}  were  introduced.
With this correction, the analytical expressions for the thermodynamics quantities obtained in the SPT can be turned 
to the CS formula \cite{Carnahan} if the length of HSC particles is set to zero and diameters of HSC and HS particle are equal.
This essentially improves the SPT description of the HS-HSC mixture at high densities.
In the present paper, we propose a modification of the CS correction, due to which the SPT expressions 
can be reduced to the Mansoori-Carnahan-Starling-Leland formula \cite{Mansoori} if the length of HSC particles is set to zero,
while the diameter HSC and HS particles can differ. The PL correction corrects the description of isotropic-nematic transition for spherocylinders of small length. 
This new formulation of the SPT description for a HS-HSC mixture is used for the reference system in the AMSA and MSA calculations.

The paper is arranged as follows. Section~2  presents the basic model and theoretical formalism. In Section~3,  the results for the
phase diagrams  are  presented and discussed.
We conclude in Section~4.

\section{Theory}

\subsection{Model}
We consider a model ionic fluid immersed in a solvent consisting of   uncharged (neutral) hard spherocylinders (HSC). The HSC  consists of a cylinder of length $L$ and diameter $D$ capped
by two hemispheres of the same diameter. The ionic fluid is treated as a
restricted primitive model (RPM) that consists of an equal number of equisized positively
and negatively charged hard spheres. The
interaction potentials between two ions are as follows:
\begin{equation}
u_{\alpha\beta}(r) = \left\{
\begin{array}{ll}
\infty, & r<\sigma_{1}\\
\displaystyle{\frac{Z_{\alpha}Z_{\beta}{\rm e}^{2}}{\varepsilon r}},& r\geqslant \sigma_{1}
\end{array}
\right. ,
\label{a1.1}
\end{equation}
where $Z_{+}=-Z_{-}=1$, $\sigma_{1}=2R_{1}$  is the  diameter of ions, $R_{1}$ is the radius of ions, ${\rm e}$ is the elementary
charge, and $\varepsilon$ is  the dielectric constant of the solvent.  We denote as $\rho_{1}=\rho_{+}+\rho_{-}$ the total number density of ions.

In the case when there is no interaction between ions and spherocylinders beyond the hard core, one can present the Helmholtz free energy of the RPM-HSC mixture in the form: 
\begin{equation}
\beta F=\beta F^{\rm{ref}}+\beta\Delta F^{\rm{ion}},
\label{F}
\end{equation}
where $F^{\rm{ref}}$ is the free energy of a reference system,  represented by a mixture of  hard-spheres and spherocylinders.
$\Delta F^{\rm{ion}}$ is the part connected
with the ionic subsystem, and $\beta=1/k_{B}T$.

\subsection{Reference system}
We start with the description of the reference system consisting of a binary mixture of hard convex bodies (HCB) of different shape. The HCB can be characterized 
by a set of three geometrical parameters:
the volume $V$ of a particle, its surface area $S$ and the mean curvature $r$ taken with a factor ${1}/{4\pi}$ \cite{holovko2015physics}.
For the HS particles with the radius $R_{1}$, these parameters are
\begin{equation}
\label{funct1}
V_{1}=\frac{4}{3}\pi R_{1}^3\;,\;\;\;   S_{1}=4\pi R_{1}^2\;,\;\;\;   r_{1}=R_{1}.
\end{equation}
For the HSC particles with the radius $R_{2}$ and the length $L_{2}$, we have
\begin{equation}
\label{funct2}
V_{2}=\pi R_{2}^2 L_{2}+\frac{4}{3}\pi R_{2}^3\;,\;\;\;   S_{2}=2\pi R_{2} L_{2}+4\pi R_{2}^2\;,\;\;\;   r_{2}=\frac{1}{4} L_{2}+R_{2}.
\end{equation}

Using the SPT theory, we can present  the equation of state  of the HS-HSC mixture in the form:
\begin{eqnarray}
\label{pressureCS}
\frac{\beta P^{\rm{ref}}}{\rho}=\frac{\beta P^{\rm{CS}}}{\rho}=\frac{\beta P^{\rm{SPT}}}{\rho}+\frac{\beta\Delta P^{\rm{CS}}}{\rho},
\end{eqnarray}
where the first addend follows from the SPT  \cite{Gibbons:1969,Boublik:1974}  
\begin{equation}
\label{pressure}
\frac{\beta P^{\rm{SPT}}}{\rho}=1+\frac{\eta}{1-\eta}+\frac{A}{2}\frac{\eta}{(1-\eta)^2}
+\frac{2B}{3}\frac{\eta^2}{(1-\eta)^3}.
\end{equation}
The second addend  is the Carnahan-Starling  (CS) correction which is introduced to improve the description of the thermodynamics of the HS-HSC mixture at high densities. The correction is chosen in the form  \cite{Boublik:1975}:
\begin{eqnarray}
\label{pressureDeltaCS}
\frac{\beta\Delta P^{\rm{CS}}}{\rho}=-\frac{\eta^3}{(1-\eta)^3}\Delta_{1}.
\end{eqnarray}  
In (\ref{pressure}), the following notations are introduced \cite{HolovkoHvozd2017}:
\begin{equation}
\label{AB}
A=\sum_{\alpha=1}^{2}x_\alpha a_\alpha, \qquad
B=\sum_{\alpha=1}^{2}x_\alpha b_\alpha,
\end{equation}
\begin{eqnarray}
a_1&=&6\frac{\eta_1}{\eta}+\left[\frac{1}{k_1}\frac{6\gamma_2}{3\gamma_2-1}
+\frac{1}{k_1^2}\frac{3(\gamma_2+1)}{3\gamma_2-1}\right]\frac{\eta_2}{\eta},
\label{a1}
\\
b_1&=&\frac{1}{2}\left(3\frac{\eta_1}{\eta}+\frac{1}{k_1}\frac{6\gamma_2}{3\gamma_2-1}\frac{\eta_2}{\eta}\right)^2,
\label{b1}
\end{eqnarray}
\begin{eqnarray}
a_2(\tau(f))&=&\left[\frac{3}{4}s_1(1+2k_1)+3k_1(1+k_1)\right]\frac{\eta_1}{\eta}
+\left[6 \right. 
\nonumber \\
&+&
\left.
\frac{6(\gamma_2-1)^2\tau(f)}{3\gamma_2-1}\right]\frac{\eta_2}{\eta}, 
\label{a2} 
\end{eqnarray}
\begin{eqnarray}
b_2(\tau(f))&=&\left(\left[\frac{3}{4}s_1+\frac{3}{2}k_1\right]\frac{\eta_1}{\eta}
+\left[\frac{3(2\gamma_2-1)}{3\gamma_2-1}+\frac{3(\gamma_2-1)^2\delta\tau(f)}{3\gamma_2-1}\right] \right.
\nonumber \\
&\times&
\left.
\frac{\eta_2}{\eta}\right)
\left(3k_1\frac{\eta_1}{\eta}+\frac{6\gamma_2}{3\gamma_1-1}\frac{\eta_2}{\eta}\right),
\label{b2}
\end{eqnarray}
where
\begin{eqnarray}
\label{k1s1gamma2}
k_1=\frac{R_2}{R_1}, \qquad s_1=\frac{L_2}{R_1}, \qquad \gamma_2=1+\frac{L_2}{2 R_2},
\end{eqnarray}
\begin{eqnarray}
\eta&=&\eta_1+\eta_2, \qquad  \eta_1=\rho_1 V_1, \qquad   \eta_2=\rho_2 V_2,
\label{eta} \\
\rho&=&\rho_1+\rho_2, \quad x_1=\frac{\rho_1}{\rho}, \qquad  x_2=\frac{\rho_2}{\rho},
\label{rho}
\end{eqnarray}
and $\delta=3/8$ is the PL correction. In Eq.(\ref{a2})-(\ref{b2}),  $\tau(f)$ is given by
\begin{equation}
\tau(f)=\frac{4}{\pi}\int f(\Omega_1)f(\Omega_2)\sin[\gamma(\Omega_1,\Omega_2)]d\Omega_1d\Omega_2,
\label{tau_f}
\end{equation}
where $\Omega=(\vartheta,\varphi)$ denotes the orientation of HSC particles and it is defined by the angles $\vartheta$ and $\varphi$, $d\Omega=\frac{1}{4\pi}\sin\vartheta d\vartheta d\varphi$ is the normalized angle element, $\gamma(\Omega_1, \Omega_2)$ is an angle between orientation vectors of two molecules, $f(\Omega)$ is the singlet orientation distribution function normalized in such a way that
\begin{equation}
\int f(\Omega)d\Omega=1.
\label{normalization}
\end{equation}
The singlet orientational distribution function $f(\Omega)$ can be obtained from a minimization of the free energy with respect to variations of this distribution. This procedure leads to  the nonlinear integral equation
\begin{equation}
\label{nonlineareq}
\ln f(\Omega_1)+\lambda+C\int f(\Omega')\sin(\Omega_1\Omega')d\Omega'=0,
\end{equation}
where
\begin{equation}
\label{constC}
C=\frac{\eta_2}{1-\eta}\left[\frac{3(\gamma_2-1)^2}{3\gamma_2-1}+
\frac{1}{1-\eta}\frac{(\gamma_2-1)^2}{3\gamma_2-1}
\delta\left(3k_1\eta_1+\frac{6\gamma_2}{3\gamma_2-1}\eta_2\right)\right]
\end{equation}
and the constant $\lambda$ is defined from the normalization condition Eq.(\ref{normalization}).

In (\ref{pressureDeltaCS}), $\Delta_{1}$ is of the form:
\begin{equation}
\Delta_{1}=\frac{q_{m}s_{m}^{2}}{9v_{m}^{2}},  
\label{Delta1}
\end{equation}
where
\begin{eqnarray}
v_{m}&=&\sum_{\alpha}x_{\alpha}V_{\alpha}, \qquad s_{m}=\sum_{\alpha}x_{\alpha}S_{\alpha}, \nonumber \\
r_{m}&=&\sum_{\alpha}x_{\alpha}r_{\alpha}, \qquad q_{m}=\sum_{\alpha}x_{\alpha}q_{\alpha},
\label{partials}
\end{eqnarray}
$q_{\alpha}=r_{\alpha}^{2}$ and the quantities $V_{\alpha}$, $S_{\alpha}$, and $r_{\alpha}$   are given in Eqs.~(\ref{funct1})-(\ref{funct2}), (\ref{rho}). 
For $\Delta_{1}=1$, the CS correction coincides with the expression used in \cite{HvozdPatsahanHolovko:2018}. In this study, we use the 
CS correction in the form  (\ref{Delta1}) under the  condition that  $R_{2}$ is the radius of  an equivalent HS with the volume which is equal 
to the volume of the HSC.  

Similar to (\ref{pressureCS}), the partial chemical potentials $\beta\mu_\alpha^{r}$ can be written as
\begin{eqnarray}
\label{chemCS}
\beta\mu_\alpha^{r}=\beta\mu_{\alpha}^{\textrm{CS}}=\beta\mu_\alpha+\beta\Delta\mu_{\alpha}^{\textrm{CS}}\;,
\end{eqnarray}
where expressions for $\beta\mu_1$ and $\beta\mu_2$ were derived in \cite{HolovkoHvozd2017}. As a result, we have  the following expressions for the chemical potential  $\beta\mu_1$
\begin{eqnarray}
\label{chem1}
\beta\mu_{1}^{\rm{SPT}}&=&\ln\Lambda_1^3\rho_1-\ln(1-\eta)
+\frac{1}{2}\frac{\eta}{1-\eta}\bigg\{a_1+6\frac{\rho_1 V_1}{\eta}+\frac{\rho_2 V_1}{\eta} \nonumber \\
&\times&
\left[\frac{3}{4}s_1(1+2k_1)+3k_1(1+k_1)\right]\bigg\}
+\frac{1}{3}\frac{\eta^2}{(1-\eta)^2}
\bigg\{b_1+3\frac{\rho_1 V_1} {\eta^2} \nonumber \\
&\times&
\left[3\eta_1+\frac{1}{k_1}\frac{6\gamma_2}{3\gamma_2-1}\eta_2\right]+
\frac{\rho_2 V_1}{\eta^2}\bigg[9k_1\left(\frac{1}{2}s_1+k_1\right)\eta_1 
\nonumber \\
&+&
\left(\frac{3}{4}\frac{6\gamma_2}{3\gamma_2-1}s_1
+3k_1\left[3+ \frac{3(\gamma_2-1)^2\delta\tau(f)}{3\gamma_2-1}\right]\right)\eta_2\bigg]\bigg\}
\nonumber \\
&+&
\beta P V_1,
\end{eqnarray}
 and for the chemical potential  $\beta\mu_2$
\begin{eqnarray}
\label{chem2}
\beta\mu_{2}^{\rm{SPT}}&=&\ln\Lambda_2^3\rho_2+\sigma(f)-\ln(1-\eta)
\nonumber \\
&+&\frac{1}{2}\frac{\eta}{1-\eta}\bigg\{a_2+\frac{\rho_1 V_2}{\eta}
\left[\frac{1}{k_1}\frac{6\gamma_2}{3\gamma_2-1}
+\frac{1}{2}\frac{1}{k_1^2}\frac{6(\gamma_2+1)}{3\gamma_2-1}\right]
\nonumber \\
&+&
\frac{\rho_2 V_2}{\eta}\left[6+\frac{6(\gamma_2-1)^2\tau(f)}{3\gamma_2-1}\right]\bigg\}
\nonumber\\
&+&\frac{1}{3}\frac{\eta^2}{(1-\eta)^2}
\bigg\{b_2+\frac{\rho_1 V_2}{\eta^2}\frac{1}{k_1}\frac{6\gamma_2}{3\gamma_2-1}
\left[3\eta_1+\frac{1}{k_1}\frac{6\gamma_2}{3\gamma_2-1}\eta_2\right]
\nonumber \\
&+&\frac{\rho_2 V_2}{\eta^2}\bigg[\left(\frac{3}{4}\frac{6\gamma_2}{3\gamma_2-1}s_1
+3k_1\left[3+\frac{3(\gamma_2-1)^2\delta\tau(f)}{3\gamma_2-1}\right]\right)\eta_1 \nonumber \\
&+&\frac{6\gamma_2}{3\gamma_2-1}\left(\frac{6(2\gamma_2-1)}{3\gamma_2-1}
+\frac{6(\gamma_2-1)^2\delta\tau(f)}{3\gamma_2-1}\right)\eta_2\bigg]\bigg\}
+\beta P V_2,
\end{eqnarray}
where the entropic term $\sigma(f)$  is defined as
\begin{equation}
\label{sigma}
\sigma(f)=\int f(\Omega)\ln f(\Omega)d\Omega
\end{equation}
and the singlet orientation distribution function $f(\Omega)$ can be obtained from Eqs.~(\ref{nonlineareq})-(\ref{constC}).	

Based on (\ref{pressureDeltaCS}) and (\ref{Delta1}), we find the expression for $\beta\Delta\mu_{\alpha}^{\textrm{CS}}$
\begin{eqnarray}
\beta\Delta\mu_{\alpha}^{\rm{CS}}&=&-\frac{V_{\alpha}}{v_{m}}\frac{\eta^{3}}{(1-\eta)^{3}}\Delta_{1}+\frac{s_{m}}{9v_{m}^{3}}\left[(q_{\alpha}s_{m}
+2S_{\alpha}q_{m})v_{m}-2V_{\alpha}q_{m}S_{m}\right] \nonumber \\
&&
\times
\left[\ln(1-\eta)+\frac{\eta}{1-\eta}-\frac{1}{2}\frac{\eta^2}{(1-\eta)^2}\right].
\label{mu_CS}
\end{eqnarray}

Putting in the above equations $L_{2}=0$, we obtain the partial chemical potential and pressure for a HS binary mixture.

\subsection{Ionic subsystem}
The contribution to free energy from the ionic subsystem will be calculated within the framework of the associative mean-spherical approximation (AMSA). 
In this case, 
$\Delta f^{\rm{ion}}=\Delta F^{\rm{ion}}/V$  can be written as \cite{Hol05,Jiang02,HolPatPat17}
\begin{equation}
\beta\Delta f^{\rm{ion}}=\beta f^{\rm{mal}}+\beta f^{\rm{el}},
\label{F_ion}
\end{equation}
where
\begin{equation}
\beta f^{\rm{mal}}=\frac{\beta F^{\rm{mal}}}{V}=\rho_{1}\ln\alpha+\frac{\rho_{1}}{2}\left(1-\alpha\right)
\label{f_mal}
\end{equation}
is the contribution from the mass action law (MAL) and
\begin{equation}
\beta f^{\rm{el}}=-\frac{\beta {\rm e}^{2}}{\varepsilon}\rho_{1}\frac{\Gamma^{B}}{1+\Gamma^{B}\sigma_{1}} +\frac{\left(\Gamma^{B} \right)^{3}}{3\pi}
\label{f_el}
\end{equation}
is the contribution from the electrostatic ion interaction. The degree of dissociation $\alpha$ satisfies the MAL
\begin{equation}
1-\alpha=\frac{\rho_{1}}{2}\alpha^{2}K,
\label{mal}
\end{equation}
where $K=K^{\gamma}K^{0}$ is the association constant with $K^{0}$ being the thermodynamic association constant.
There is a certain kind of arbitrariness in how we define the ion pair and hence the thermodynamic association constant. Here, $K^{0}$ is chosen in the form proposed in  \cite{Olaussen91}. The corresponding association constant $K^{0}$ gives the best estimations
for the vapour-liquid critical parameters obtained within the framework of the ionic association approach for the bulk RPM \cite{Jiang02}. 
In the AMSA, $K^{\gamma}$ is given by \cite{Blum95,Bernard96}
\begin{equation}
K^{\gamma}=g_{11}(\sigma_{1})\exp\left[-b\frac{\Gamma^{B}\sigma_{1} (2+\Gamma^{B}\sigma_{1})}
{(1+\Gamma^{B}\sigma_{1})^{2}}\right],
\label{K_gamma}
\end{equation}
where $b=\lambda_{B}/\sigma_{1}=\beta {\rm e}^{2}/\sigma_{1}\varepsilon$ is
the dimensionless Bjerrum length,  $\Gamma^{B}$ is the screening parameter calculated from the equation 
\begin{equation}
4\left(\Gamma^{B}\right)^{2}\left(1+\Gamma^{B}\sigma_{1}\right)^{3}=\kappa_{D}^{2}\left(\alpha+\Gamma^{B}\sigma_{1}\right),
\label{Gamma_B}
\end{equation}
where  $\kappa_{D}^{2}=4\pi q^{2}\rho_{1}/(\varepsilon k_{B}T)$ is the inverse squared Debye length.
It should be noted that without association ($\alpha=1$), $\Gamma^{B}$
reduces to the screening parameter $\Gamma$ in the MSA \cite{Waisman72,WaismanLeb72,Blum74,Blum75}
\begin{equation}
\Gamma\sigma_{1}=\frac{1}{2}[\sqrt{1+2\kappa_{D}\sigma_{1}}-1].
\label{Gamma}
\end{equation}
In (\ref{K_gamma}), $g_{11}(\sigma_{1})$ is the contact value of the
radial distribution function of ions at presence of the solvent molecules.  For the solvent molecules  modelled by  spherocylinders,  $g_{11}(\sigma_{1})$ is of the form:
\begin{eqnarray}
g_{11}(\sigma_{1})&=&\frac{1}{1-\eta}+\frac{3}{2}\frac{1}{(1-\eta)^2}
\left(\eta_1+\frac{1}{k_1}\frac{2\gamma_2}{3\gamma_2-1}\eta_2\right) \nonumber \\
&+&
\frac{1}{2}\frac{1}{(1-\eta)^3}\left(\eta_1+\frac{1}{k_1}\frac{2\gamma_2}{3\gamma_2-1}\eta_2\right)^2,
\label{g11}
\end{eqnarray}
where $\eta=\eta_{1}+\eta_{2}$ is the total volume fraction of ions and solvent molecules, $k_{1}$ and $\gamma_2$ are given in (\ref{k1s1gamma2}).


Using  Eqs. (\ref{f_mal}) and (\ref{mal}), one obtains the following expressions for $P^{\rm{mal}}$ and $\mu_{1}^{\rm{mal}}$:
\begin{eqnarray}
\beta P^{\rm{mal}}&=&-\frac{\rho_1}{2}(1-\alpha)\left(1+\rho_1\frac{\partial\ln K^{\gamma}}{\partial\rho_1}\right),
\label{P_mal} \\
\beta\mu^{\rm{mal}}_{1}&=&\ln\alpha-\frac{\rho_1}{2}(1-\alpha)\frac{\partial\ln K^{\gamma}}{\partial\rho_1}.
\label{mu_mal}
\end{eqnarray}

For the electrostatic contribution (\ref{f_el}), we use the simple interpolation scheme known as the SIS approximation 
\cite{Stell89}. Within the framework of this approach, the effects
of ionic screening are described accurately, but the effects of ionic association are neglected.  As a result, one obtains
\begin{equation}
\beta P^{\rm{el}}=-\frac{\Gamma^{3}}{3\pi}, \qquad
\beta\mu_{1}^{\rm{el}}=-\frac{1}{T^{*}}\frac{\Gamma\sigma_1}{(1+\Gamma\sigma_1)},
\label{P_mu_el}
\end{equation}
where  
 $\Gamma$ is given in (\ref{Gamma}) and  $T^{*}=b^{-1}$. 

Taking into account Eqs.~(\ref{pressureCS})-(\ref{sigma}), (\ref{P_mal})-(\ref{mu_mal}), and (\ref{P_mu_el}), we  present   pressure  and  the partial  chemical potentials of our system   as follows:
\begin{eqnarray}
\beta P^{\rm{AMSA}}&=&\beta P^{\rm{\rm{CS}}}+\beta P^{\rm{mal}}+\beta P^{\rm{el}},
\label{P_amsa} \\
\beta\mu_{1}^{\rm{AMSA}}&=&\beta\mu_{1}^{\rm{CS}}+\beta\mu^{\rm{mal}}_{1}+
\beta\mu^{\rm{el}}_{1},
\label{nu_i_amsa} \\
\beta\mu_{s}^{\rm{AMSA}}&=&\beta\mu_{s}^{\rm{CS}}.
\label{nu_s_amsa}
\end{eqnarray}
These equations are used below for the calculation of  phase diagrams. In our calculations, we supplement the AMSA with the SIS
\cite{Stell89}  substituting  $\Gamma$ (see (\ref{Gamma})) for $\Gamma^{B}$ in Eqs.~(\ref{P_mu_el}).  
This approximation is equivalent to the Wertheim first-order thermodynamic perturbation theory \cite{Wertheim84}.
The approach  gives  a reasonable
estimate  for the location of the critical point of the  RPM fluid. 

Neglecting in   (\ref{P_amsa})-(\ref{nu_i_amsa}) the addends
connected with associations one arrives at the pressure and the chemical potentials in the MSA.  

\section{Results and Discussion}
We consider two models -- a binary mixture of RPM and HSC particles (RPM-HSC) 
and a binary mixture of the RPM and HS particles (RPM-HS). 
In the latter model, the volume of a solvent particle (neutral HS) is taken of the same value as the one of a hard 
spherocylinders (neutral HSC) in the former model. 
According to the condition of equality of the HSC and HS volumes ($V_{2}^{HS}=V_{2}^{HSC}$) and using (\ref{funct1})-(\ref{funct2})
the following relationship between the diameters of HS and HSC solvent particles ($\sigma_{2}=2R_{2}$) 
can be derived in the following form:
\begin{eqnarray}
\sigma_{2}^{{\rm HS}}=\sigma_{2}^{{\rm HSC}}[(3\gamma_{2}-1)/2]^{1/3},
\label{sigma_ekv}
\end{eqnarray}
where $\gamma_{2}$ is given in (\ref{k1s1gamma2}). In this study, we consider spherocylinders with $L_{2}=5\sigma_{1}$ (where $\sigma_{1}=2R_{1}$), 
then  the diameter of an equivalent HS is equal to $2.0408\sigma_{1}$.

In this section we present results for the phase diagrams obtained from both the MSA and the AMSA. Coexistence curves are calculated at subcritical 
temperatures using the conditions of two-phase equilibrium
\begin{eqnarray}
\mu_{i}(\rho^{\alpha},x_{1}^{\alpha},T)&=&\mu_{i}(\rho^{\beta},x_{1}^{\beta},T), \label{nui_eq} \\ 
\mu_{s}(\rho^{\alpha},x_{1}^{\alpha},T)&=&\mu_{s}(\rho^{\beta},x_{1}^{\beta},T), \label{nus_eq} \\ 
P(\rho^{\alpha},x_{1}^{\alpha},T)&=&P(\rho^{\beta},x_{1}^{\beta},T), 
\label{P_eq}
\end{eqnarray}
 where $\rho^{\alpha(\beta)}$ is the total number density ($\rho=\rho_{1}+\rho_{2}$) in phase $\alpha(\beta)$ and $x_{1}^{\alpha(\beta)}$ is the concentration in 
 phase $\alpha(\beta)$ of ions, $x_{1}=\rho_{1}/\rho$.
The phase diagrams are built by solving numerically a set of equations Eqs.~(\ref{nui_eq})-(\ref{P_eq}) with respect to the densities $\rho^{\alpha}$ and 
$\rho^{\beta}$ and one of the concentrations $x_{1}^{\alpha}$ when the second concentration $x_{1}^{\beta}$
is given. Therefore, a series of the densities and concentrations in phases $\alpha$ and $\beta$ are obtained at temperatures of wide range.
To solve the set of equations Eqs.~(\ref{nui_eq})-(\ref{P_eq}), the Newton-Raphson iterative procedure has been used with an accuracy $10^{-9}$. 
In the following, length is measured in $\sigma_{1}$ units and 
we introduce   dimensionless units for the temperature, pressure, and  total number density
\begin{eqnarray}
T^{*}=k_{B}T\varepsilon\sigma_{1}/{\rm e}^{2}, 
\qquad
P^{*}=P\varepsilon\sigma_{1}^{4}/{\rm e}^{2}, \qquad
\rho^{*}=\rho\sigma_{1}^{3}.
\label{notation}
\end{eqnarray}

\begin{figure}[htbp]
	\begin{center}
		\includegraphics[width=0.47\textwidth,angle=0,clip=true]{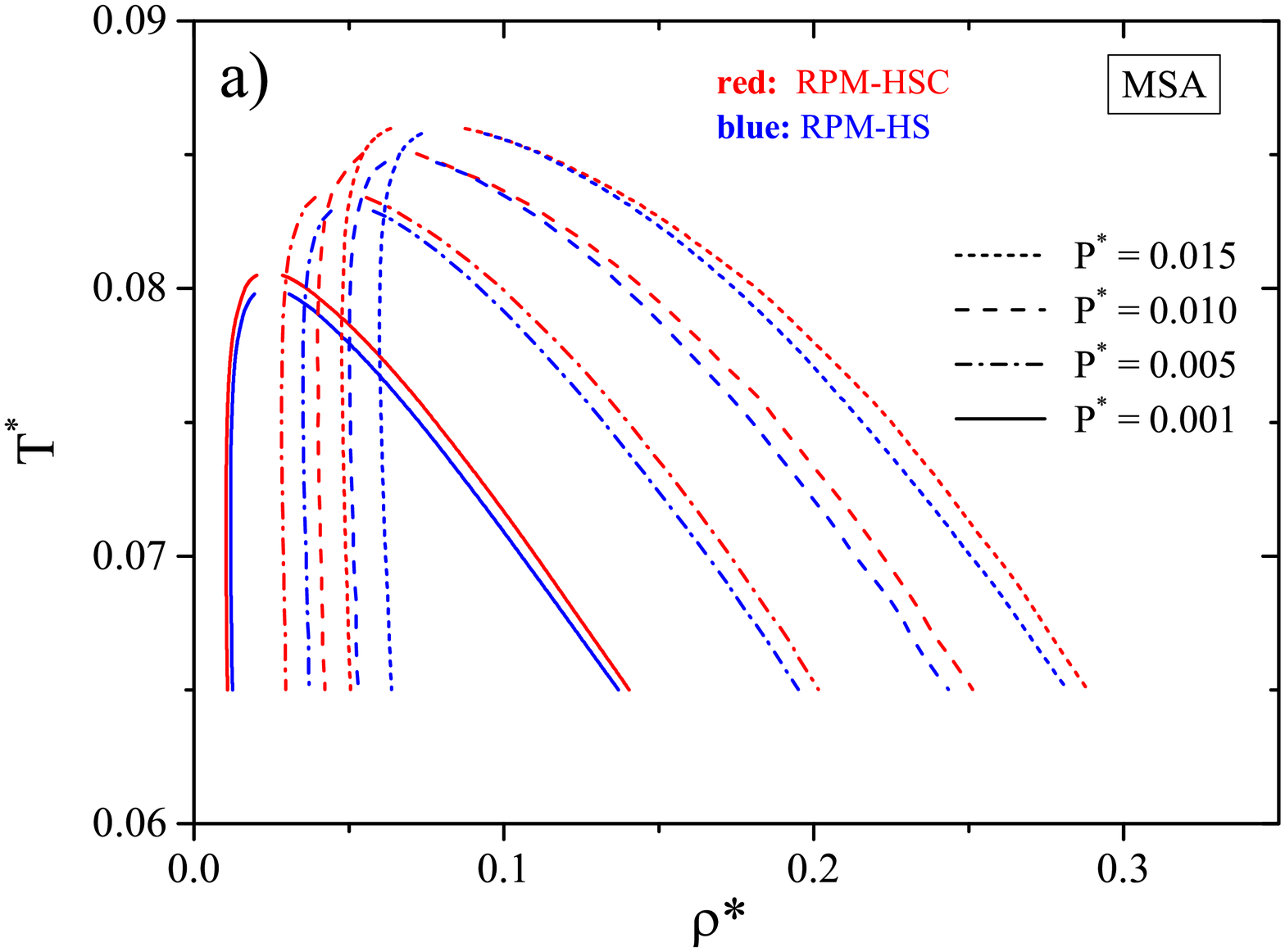}
		\includegraphics[width=0.47\textwidth,angle=0,clip=true]{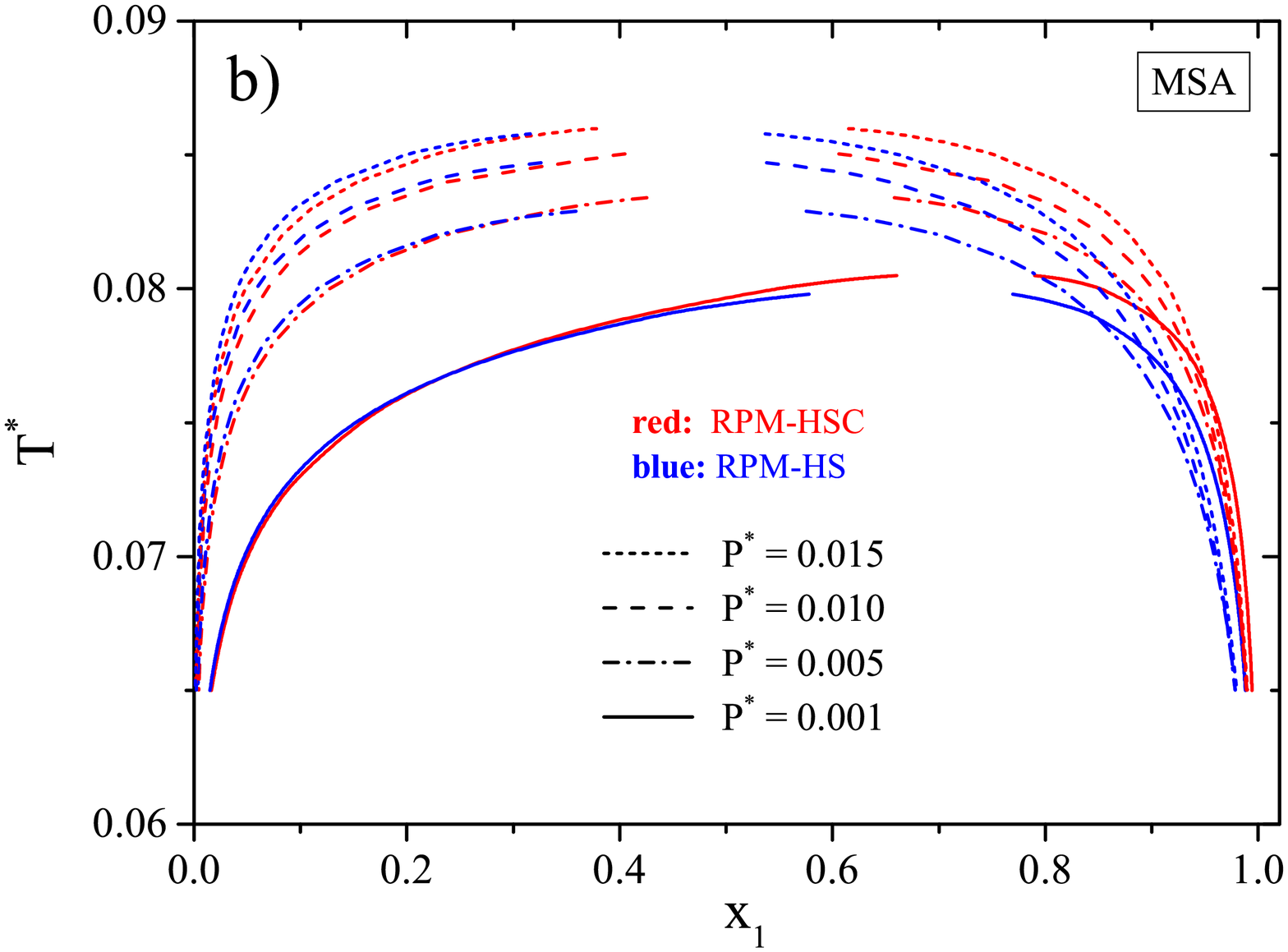}
		\caption{\label{fig:coexMSA} Coexistence curves of the RPM/HSC and RPM/HS mixtures in $T^{*}$-$\rho^{*}$ (a) and $T^{*}$-$x_{1}$ (b) planes at 
		constant reduced pressures in the MSA approximation. $T^{*}$, $P^{*}$, and $\rho^{*}$ are defined in Eq.~(\ref{notation}), and $x_{1}=\rho_{1}/\rho$.}
	\end{center}
\end{figure}
First, we calculate the phase diagrams in the MSA. In this case, we neglect in Eqs.~(\ref{P_amsa})-(\ref{nu_i_amsa}) 
the contributions connected with the MAL.
In Figs.~\ref{fig:coexMSA}~(a)-(b), we show the coexistence curves which are calculated at the selected pressures 
using Eqs.~(\ref{nui_eq})-(\ref{P_eq}) and presented in the $T^{*}$-$\rho^{*}$ and $T^{*}$-$x_{1}$ planes ($x_{1}=\rho_{1}/\rho$ 
is the concentration of ions). The considered pressures are higher than the critical pressure of the RPM in the MSA ($P^{*}=9.64\times 10^{-5}$). 
It is worth noting that the MSA predictions for the critical temperature and the critical 
density of the RPM are as follows  \cite{Cai-Mol1}: $T_{c}^{*}=0.07858$ and $\rho_{c}^{*}=0.01449$. As one can see from Fig.~\ref{fig:coexMSA}~(a),
an increase of pressure shifts the coexistence region towards higher total number densities and towards higher reduced temperatures. Simultaneously, 
the width of the coexistence region becomes larger. This behaviour is observed for the both models, RPM-HSC and RPM-HS. 
However, the coexistence envelopes in the case of the RPM-HSC mixture (red curves) are broader than those in the case of RPM-HS mixture (blue curves).  
Broadening of the coexistence region with an increase of pressure is also observed for the both models 
in the $T^{*}$-$x_{1}$ plane (Fig.~\ref{fig:coexMSA}~(b)). As in  Fig.~\ref{fig:coexMSA}~(a), the coexistence envelopes of the RPM-HSC mixture (red curves)
are wider than those of the RPM-HS mixture (see Fig.~\ref{fig:coexMSA}~(b)). At the same time, this difference is more essential for the ion rich branches than for the ion poor branches. One can also notice that the critical concentration of ions decreases with increasing pressure.

\begin{figure}[htbp]
	\begin{center}
		\includegraphics[width=0.47\textwidth,angle=0,clip=true]{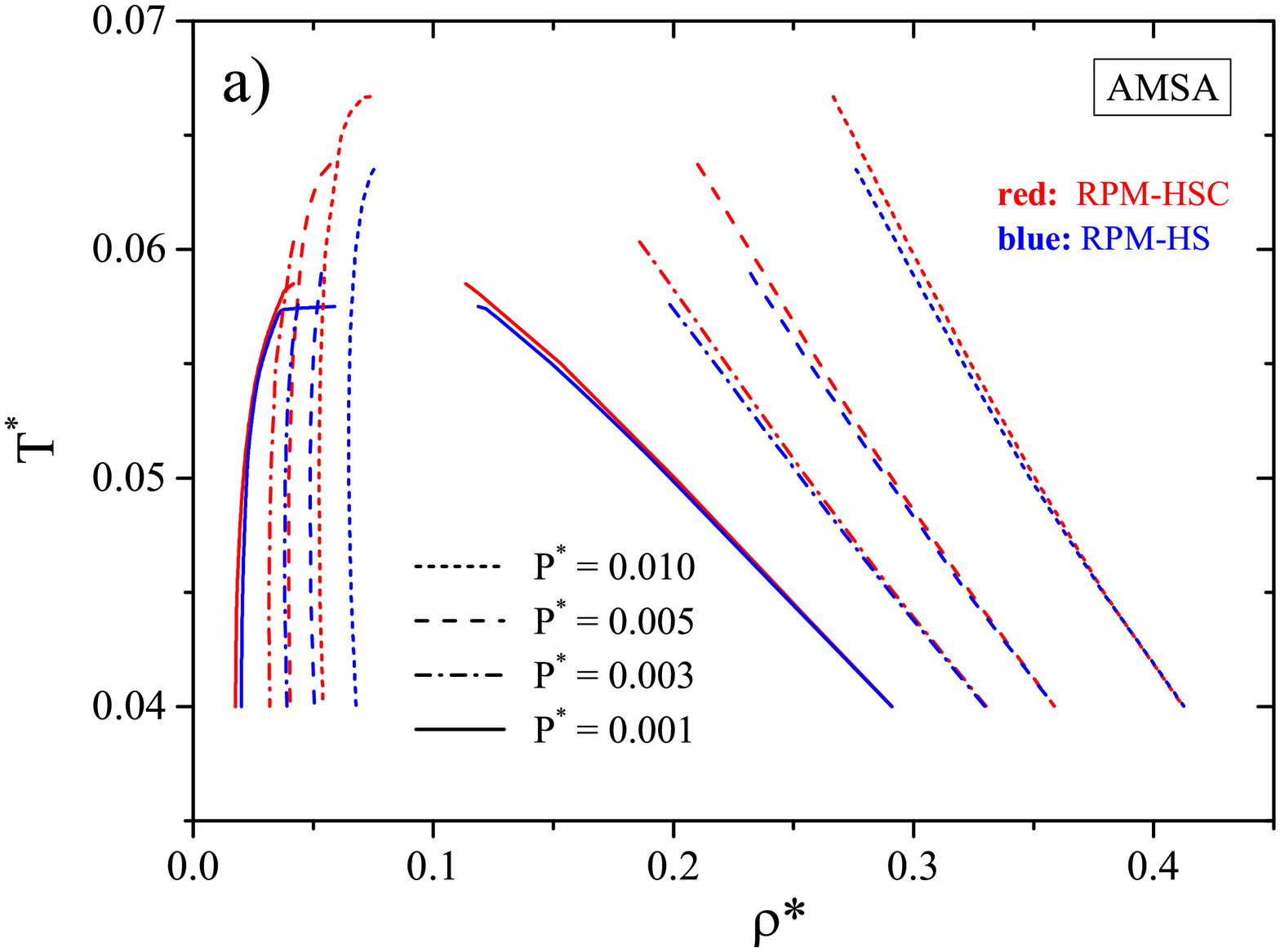}
		\includegraphics[width=0.47\textwidth,angle=0,clip=true]{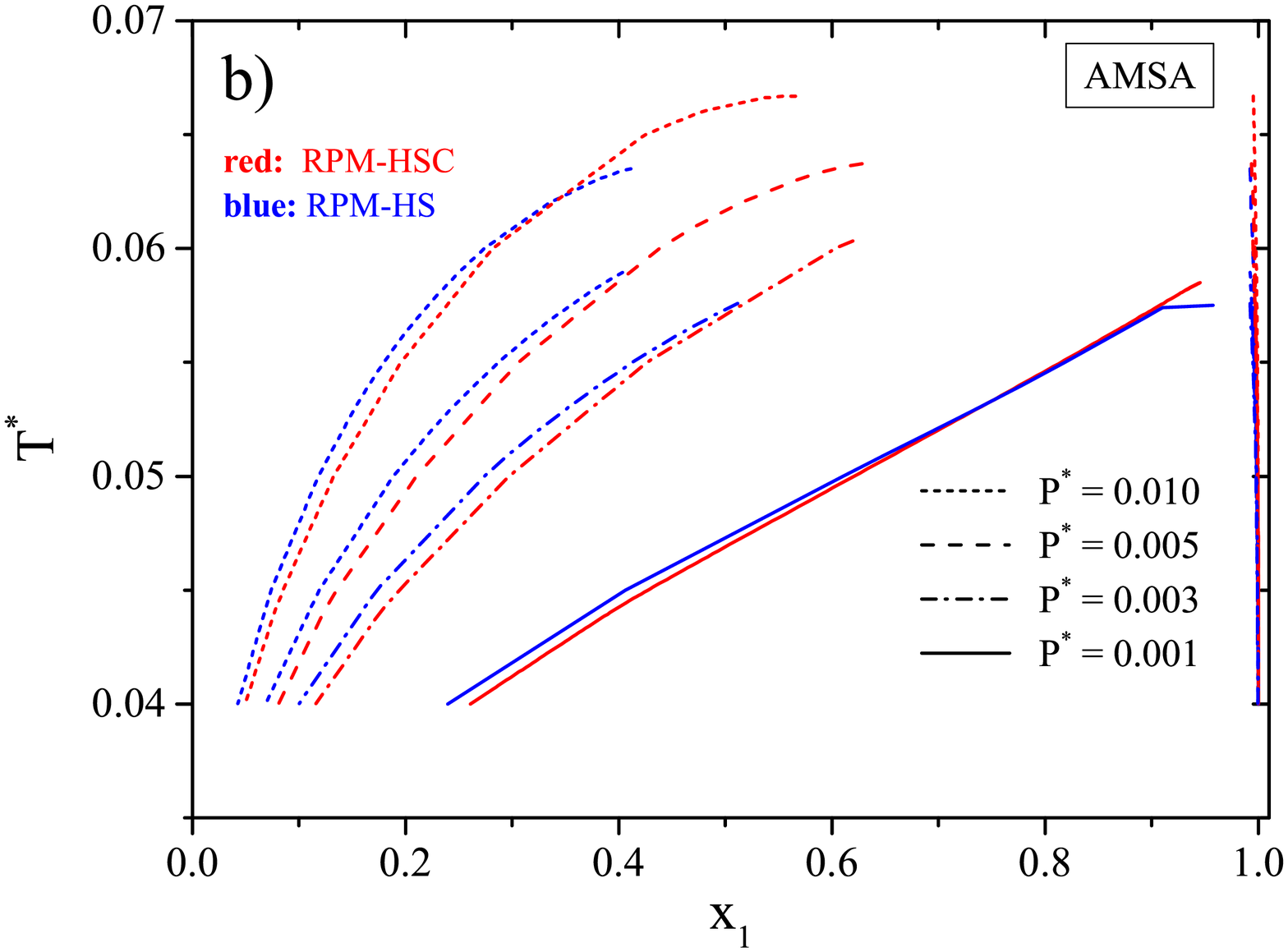}
		\caption{\label{fig:coexAMSA} Coexistence curves of the RPM/HSC and RPM/HS mixtures in  $T^{*}$-$\rho^{*}$ (a) and $T^{*}$-$x_{1}$ (b) 
		planes at constant reduced pressures in the AMSA approximation. $T^{*}$, $P^{*}$, and $\rho^{*}$ are defined in Eq.~(\ref{notation}), and 
		$x_{1}=\rho_{1}/\rho$.}
	\end{center}
\end{figure}

Now, let us consider the phase diagrams obtained from the AMSA theory. In this case, the partial chemical potentials and pressure 
are given by 
Eqs.~(\ref{P_mal})-(\ref{nu_s_amsa}) supplemented by the solution of Eq.~(\ref{Gamma_B}). 
The association constant is  chosen to be  $K^{0}=12K_{Eb}^{0}$ ($K^{0}_{Eb}$ is the association constant introduced by
Ebeling \cite{Ebeling68,HolPatPat17}).  As in the MSA, the  selected  
values of pressure are higher than the critical pressure of the RPM. 
The reduced critical parameters of the RPM calculated in the AMSA are as follows \cite{Jiang02}:
$T_{c}^{*}=0.0492$, $P_{c}^{*}=7.44\times 10^{-4}$, and $\rho_{c}^{*}=0.059$. It should be noted that the critical pressure of the RPM obtained 
from the AMSA theory is by an order higher than the corresponding pressure found from the MSA. Accordingly,  for the RPM, the AMSA  critical density 
is by four times higher than the critical density obtained from the MSA. 

In Figs.~\ref{fig:coexAMSA}~(a)-(b), 
we present the coexistence curves calculated using the AMSA theory. In general, the trends of the ($T^{*}$,$\rho^{*}$) phase diagrams (Fig.~\ref{fig:coexAMSA}~(a)) 
with an increase of pressure 
are similar to the corresponding trends found in the MSA, although some deviations are observed. In the AMSA, the difference between the high-density branches 
obtained for the RPM-HSC and RPM-HS mixtures slightly increases with an increase of temperature (Fig.~\ref{fig:coexAMSA}~(a)) while an opposite trend is found in the MSA (see Fig.~\ref{fig:coexMSA}~(a)). 
 Different trends have been also noticed in the critical region. Nevertheless, in the most cases the critical temperature in the RPM-HSC system 
is higher than in the RPM-HS system, and one can see that at the higher pressures it becomes almost identical for the both models within the MSA, 
while in the AMSA the difference between critical temperatures of the RPM-HSC and RPM-HS remains essential at all considered pressures.
The differences between the results obtained in the MSA and AMSA appear due to the MAL term in the AMSA theory, which takes into account the associative
interaction caused by strong attraction between positively and negatively charged ions of RPM. Moreover, the MAL contribution
depends on the contact value  between ionic particles (\ref{g11}), and it strongly depends on the densities and concentrations in the system.
At same time, the strength of association is also defined by the temperature.

Regarding the ($T^{*}$,$x_{1}^{*}$) diagrams obtained in the AMSA (Fig.~\ref{fig:coexAMSA}~(b)), they substantially differ from
those shown in Fig.~\ref{fig:coexMSA}~(b). Contrary to the MSA results, the phase diagrams obtained for the both models from the AMSA show 
that the high-density phase mostly consists of ions at all the  considered values of pressure. Therefore,  the 
effects of association between ions being taken into account can lead to the phase diagrams which differ qualitatively from the phase diagrams calculated in the MSA.

It should be noted that we have not observed any isotropic-nematic phase transition in the RPM-HSC system, i.e. in all our calculations
HSC particles have been in the isotropic state.
This is because at the given parameters, the density and concentration of HSC particles 
do not reach appropriate values at which the nematic phase can form.
For instance, it is known \cite{HvozdPatsahanHolovko:2018} (case $L_{2}=5.0\sigma_2$) that for the pure HSC system ($x_{1}=0$), a stable nematic phase appears at the HSC density around $0.094$. One can see that in the ion poor regions, where $x_{1}$ is rather small, the density of HSC particles is far lower than $0.094$. Furthermore, it was found from the SPT for the binary mixture of uncharged HS and HSC particles \cite{HvozdPatsahanHolovko:2018} that
the nematic phase cannot be formed at all if the concentration of HS particles is higher than $0.4$. 
As it is seen from our diagrams, the concentration of ions $x_{1}$ is always higher than $0.4$, where the density $\rho^{*}$ is larger than $0.094$.
However, we cannot exclude a possibility of the isotropic-nematic phase transition in RPM-HSC systems with an increasing pressure, where
the particles should be packed denser, or for the longer HSC particles, where the nematic phase can exist at lower densities.
Moreover, in this case one can obtain more than two coexisting phases simultaneously, i.e., apart from the lower density isotropic phase
one can observe a higher density isotropic phase and a  higher density nematic phase. This point needs a more comprehensive study.

\section{Conclusion}
We have studied the fluid-fluid phase behaviour of the explicit solvent model represented as a binary mixture of oppositely charged HS 
of equal diameters (RPM) and neutral hard spherocylinders (HSC). 
To describe this model, we have proposed the approach combining the SPT theory and the AMSA approximation, 
and on this basis we have derived the expressions for the partial chemical potential and pressure. More specifically, the thermodynamic properties 
of the reference system represented as a binary mixture of neutral HS and neutral HSC has been obtained using the SPT approach. 
For an ionic subsystem we have used the SIS-AMSA theory, which corresponds to the Wertheim first-order thermodynamic perturbation theory. 
For comparison, the ionic subsystem has been also treated within the MSA theory.

We have considered two models: the RPM-HSC mixture and an ``equivalent'' RPM-HS mixture. 
In the  former case, we have restricted our attention to the HSC with an aspect ratio $L_{2}/\sigma_{2}=5$ and RPM particles of diameter $\sigma_{1}=\sigma_{2}$. 
In the latter model, the solvent is represented as HS particles with the volume equal to the volume of the HSC particles of the former case.  
It is known that for totally uncharged mixtures of HSC and HS, at low densities, HSC component exists in isotropic state, i.e., no orientation order is expected. At sufficiently high densities, a HSC component can form the nematic phase, and due to a subtle balance between entropic contributions 
from  different components, the demixing phenomena can take place, where the HSC and HS start to redistribute between nematic and isotropic phases~\cite{HvozdPatsahanHolovko:2018,cuetos2007,lago2004}. 
We have not found the nematic phase in the RPM-HSC model at pressures considered in the present study. 
However, our preliminary results suggest a possibility of isotropic-nematic phase transitions at higher pressures or/and longer HSC particles.
This requires a more thorough investigation, which will be done in future studies.

We have calculated the fluid-fluid coexistence curves of the RPM-HSC  and RPM-HS mixtures using the equations of phase equilibrium. The analysis of 
the results have demonstrated that both models show a similar fluid-fluid phase behaviour although quantitative difference exists.  
It is seen from the MSA phase diagrams that the critical temperature increases with an increase of pressure and this dependence is more
noticeable than in the case of an equisized RPM-HS mixture (see \cite{Patsahan-Patsahan:2018}). 
Although we have not managed to get the critical points in the AMSA, it is clearly seen that the coexistence envelopes of the RPM-HSC 
and RPM-HS mixtures obtained from the both theories shift towards higher total number densities and towards higher reduced temperatures when the pressure increases. Simultaneously, the coexistence envelopes become broader. At the same time, the AMSA and 
MSA phase diagrams of the both models presented in the temperature-concentration plane considerably differ. The AMSA phase diagrams show that 
the high-density phase mostly consists of the ions for all pressures considered in this study. This result also differs from the results recently obtained for the RPM-HS mixture with $\sigma_{1}=\sigma_{2}$ \cite{Patsahan-Patsahan:2018} where similar situation was observed only for low pressures. 

To the best of our knowledge, we have made the first attempt to theoretically describe  the fluid-fluid equilibrium in an explicit solvent model 
in which the solvent molecules are of non-spherical shape. We have shown the effect of asphericity of solvent molecules on  the phase behavior
in such systems.  It was shown that at the given parameters, the considered RPM-HSC model was in isotropic state.
However, the effect of asphericity of solvent molecules can be more essential if the fluid-fluid phase transition is accompanied by
the isotropic-nematic phase transitions. Hence,  we suggest to consider the higher pressures and longer spherocylinder particles.
 It is also important to take into account attractive interactions between solvent particles, or/and an attraction between solvent molecules
and ions.


\section*{Acknowledgement}
This work was partly supported by the European Union's Horizon 2020 research and
innovation programme under the Marie Sk{\l}odowska-Curie grant agreement No 734276.

\end{document}